\documentclass[sigconf]{acmart}
\AtBeginDocument{%
  }

\copyrightyear{2026}
\acmYear{2026}
\setcopyright{cc}
\setcctype{by}
\acmConference[CHI EA '26]{Extended Abstracts of the 2026 CHI Conference on Human Factors in Computing Systems}{April 13--17, 2026}{Barcelona, Spain}
\acmBooktitle{Extended Abstracts of the 2026 CHI Conference on Human Factors in Computing Systems (CHI EA '26), April 13--17, 2026, Barcelona, Spain}
\acmDOI{10.1145/3772363.3798474}
\acmISBN{979-8-4007-2281-3/2026/04}

\usepackage{subcaption}

\definecolor{hotpink}{RGB}{255, 83, 115}

\begin{document}

\title{Break the Window: Exploring Spatial Decomposition of Webpages in XR}


\author{Chenyang Zhang}
\orcid{0009-0003-1116-4895}
\affiliation{
  \institution{Georgia Institute of Technology}
  \city{Atlanta}
  \state{Georgia}
  \country{USA}
}
\email{chenyang.zhang@gatech.edu}

\author{Tianjian Wei}
\orcid{0009-0000-9368-9756}
\affiliation{
  \institution{Georgia Institute of Technology}
  \city{Atlanta}
  \state{Georgia}
  \country{USA}
}
\email{twei79@gatech.edu}

\author{Haoyang Yang}
\orcid{0000-0002-0566-0169}
\affiliation{
  \institution{Georgia Institute of Technology}
  \city{Atlanta}
  \state{Georgia}
  \country{USA}
}
\email{alexanderyang@gatech.edu}

\author{Mar Gonzalez-Franco}
\orcid{0000-0001-6165-4495}
\affiliation{
  \institution{Google}
  \city{Seattle}
  \state{Washington}
  \country{USA}
}
\email{margon@google.com}

\author{Yalong Yang}
\authornote{These authors contributed equally to this work and share co-senior authorship.}
\orcid{0000-0001-9414-9911}
\affiliation{
  \institution{Georgia Institute of Technology}
  \city{Atlanta}
  \state{Georgia}
  \country{USA}
}
\email{yalong.yang@gatech.edu}

\author{Eric J Gonzalez}
\authornotemark[1]
\orcid{0000-0002-2846-7687}
\affiliation{
  \institution{Google}
  \city{Seattle}
  \state{Washington}
  \country{USA}
}
\email{ejgonz@google.com}

\renewcommand{\shortauthors}{Zhang et al.}

\begin{abstract}
Most XR web browsers still present webpages as a single floating window, carrying over desktop design assumptions into immersive space. We explore an alternative by \emph{breaking the browser window} and distributing a webpage into spatial UI chunks within a mixed-reality workspace. We present \emph{Break-the-Window} (BTW), an exploratory prototype that spatially decomposes live, fully functional webpages into movable panels supporting mid-air and surface-attached placement, as well as direct touch and ray-based interaction. Through a formative study with XR practitioners and an exploratory qualitative study with 15 participants, we observed how spatial decomposition supports distributed attention and spatial meaning-making, while also surfacing challenges around coordination effort, interaction precision, and the lack of shared spatial UI conventions. This work invites discussion on how web interfaces might be reimagined for spatial computing beyond the single-window paradigm.
\end{abstract}



\begin{teaserfigure}
  \setlength{\abovecaptionskip}{4pt} 
  \includegraphics[width=\textwidth]{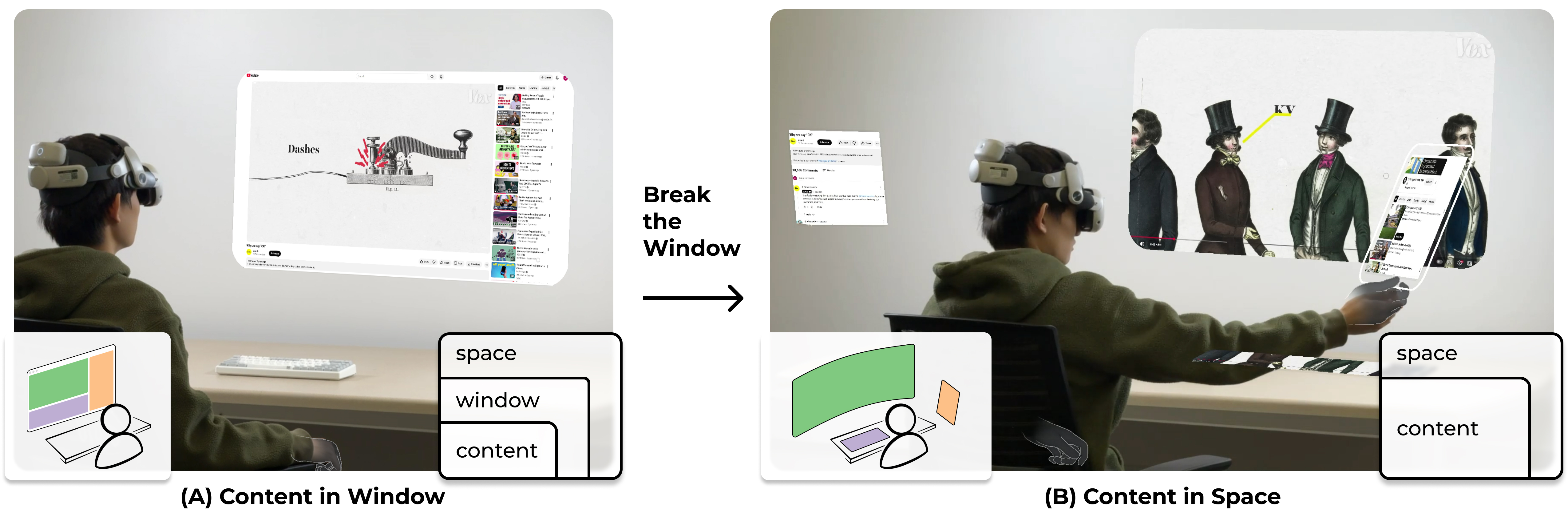}
    \caption{
    \textbf{Break-the-Window (BTW).}
    Conventional XR web experiences place a complete 2D browser window into 3D space, preserving a \emph{content $\rightarrow$ window $\rightarrow$ space} design hierarchy in which web content remains constrained by the window and cannot fully leverage spatial properties (left).
    BTW rethinks this relationship by breaking the window itself and decomposing webpages into spatially distributed UI chunks, forming a \emph{content $\rightarrow$ space} design hierarchy (right).
    This paradigm shift enables users to engage with web information more directly in space, supporting spatial organization and embodied interaction, and fostering a direct dialogue between the structural semantics of webpages and the spatial semantics of the surrounding environment.
    }
  \Description{}
  \label{fig:teaser}
\end{teaserfigure} 


\maketitle


\section{Introduction}

Web browsing remains one of the most common digital activities, yet today's XR web experiences largely reproduce a legacy interaction model: a single 2D browser window floating in 3D space. This window-centric paradigm, rooted in the PC/WIMP tradition~\cite{jacob2008reality}, preserves a \emph{content $\rightarrow$ window $\rightarrow$ space} hierarchy and limits how users can leverage spatial organization, embodied interaction, and spatial memory when working with multi-faceted web content.

Emerging spatial computing platforms enable information to be distributed throughout the surrounding environment and accessed through embodied input such as gesture, gaze, and ray-based interaction~\cite{hertel2021taxonomy}. Prior work has explored spatial presentation of media and information in XR~\cite{laviola20173d, lu2025ego} and adaptive placement strategies based on relevance or context~\cite{cheng2021semanticadapt, luo2022should}. However, general web browsing in XR remains largely bound to the single-window metaphor, with entire webpages placed into space as flat surfaces rather than rethought as spatial structures.

In this work, we ask: \textbf{What if we break the browser window itself?}
Instead of treating a webpage as an indivisible surface, we explore decomposing it into functional UI chunks and distributing these chunks across a spatial workspace, allowing the structural semantics already embedded in webpages (e.g., primary content, controls, peripheral information) to directly engage with the spatial semantics of XR (Figure~\ref{fig:teaser}).

We present \textbf{\emph{Break-the-Window} (BTW}), a positional research prototype that operationalizes this idea. BTW decomposes fully functional webpages into multiple spatial panels that users can place, move, resize, and reorganize within a desk-centered mixed-reality workspace. The system supports both mid-air and surface-attached layouts, as well as direct touch and ray-based interaction, while preserving existing webpage functionality without redesigning or modifying the underlying web interfaces.

To ground this exploration, we conducted a formative study with six XR practitioners to examine how experts conceptualize webpage decomposition, spatial roles, and placement--interaction relationships. We then ran a qualitative study with 15 participants, where BTW was experienced alongside desktop browsing and a single-window XR browser as reference points. Our findings show that breaking the window shifts browsing from interacting with a single view toward inhabiting a blended physical--digital workspace: attention and interaction become distributed in space, layout becomes a semantic resource, and physical anchoring both stabilizes and complicates interaction. At the same time, participants surfaced frictions around coordination cost, embodied precision, productivity trade-offs, and the lack of shared spatial UI conventions.

Together, this poster uses BTW as a concrete design prototype to invite discussion on what a \emph{spatial UI grammar} for web browsing could look like beyond the single-window paradigm.

\section{Related Work}

Prior work has explored redistributing interface elements across devices and spatial environments. AR-enhanced widgets and mobile-to-AR techniques extend smartphone content into surrounding space to support precision, comparison, and bookmarking~\cite{brasier2021ar, wieland2024push2ar}. Cross-device systems such as XDBrowser~\cite{nebeling2016xdbrowser} and WinCuts~\cite{tan2004wincuts} demonstrate how users decompose and replicate interface regions across screens, revealing benefits of fine-grained fragmentation and synchronization. In immersive contexts, recent work has examined how relations between multiple views can be explicitly represented in MR, synthesizing a design space of visual association techniques to support coordination and sensemaking across spatially distributed views~\cite{luo2026beyond}. Together, these approaches highlight the value of breaking monolithic interfaces, but primarily emphasize distribution and association rather than restructuring web content itself within XR.

Research in spatial computing further positions space as an organizational and semantic resource. Systems such as Spatialstrates~\cite{borowski2025spatialstrates} and InteractionAdapt~\cite{cheng2023interactionadapt} support composable cross-reality workspaces and layout adaptation based on physical affordances. Studies of large displays and multi-monitor environments show that physical navigation and spatial partitioning enhance performance and peripheral awareness~\cite{ball2007move, grudin2001partitioning}, and industry guidelines advocate distributing content into spatial panels in XR~\cite{androidxr2025spatialui}. Commercial systems such as Apple's visionOS introduce spatial browsing that extends webpages into immersive presentations while largely preserving the page as a unified surface~\cite{apple2025visionos26}. Building on these directions, BTW investigates decomposing functional webpages into spatially arranged UI chunks and examines how these spatialized components acquire new semantic meaning through their placement, relation, and persistence in a mixed-reality workspace.

\begin{figure}
  \centering
  \includegraphics[width=\linewidth]{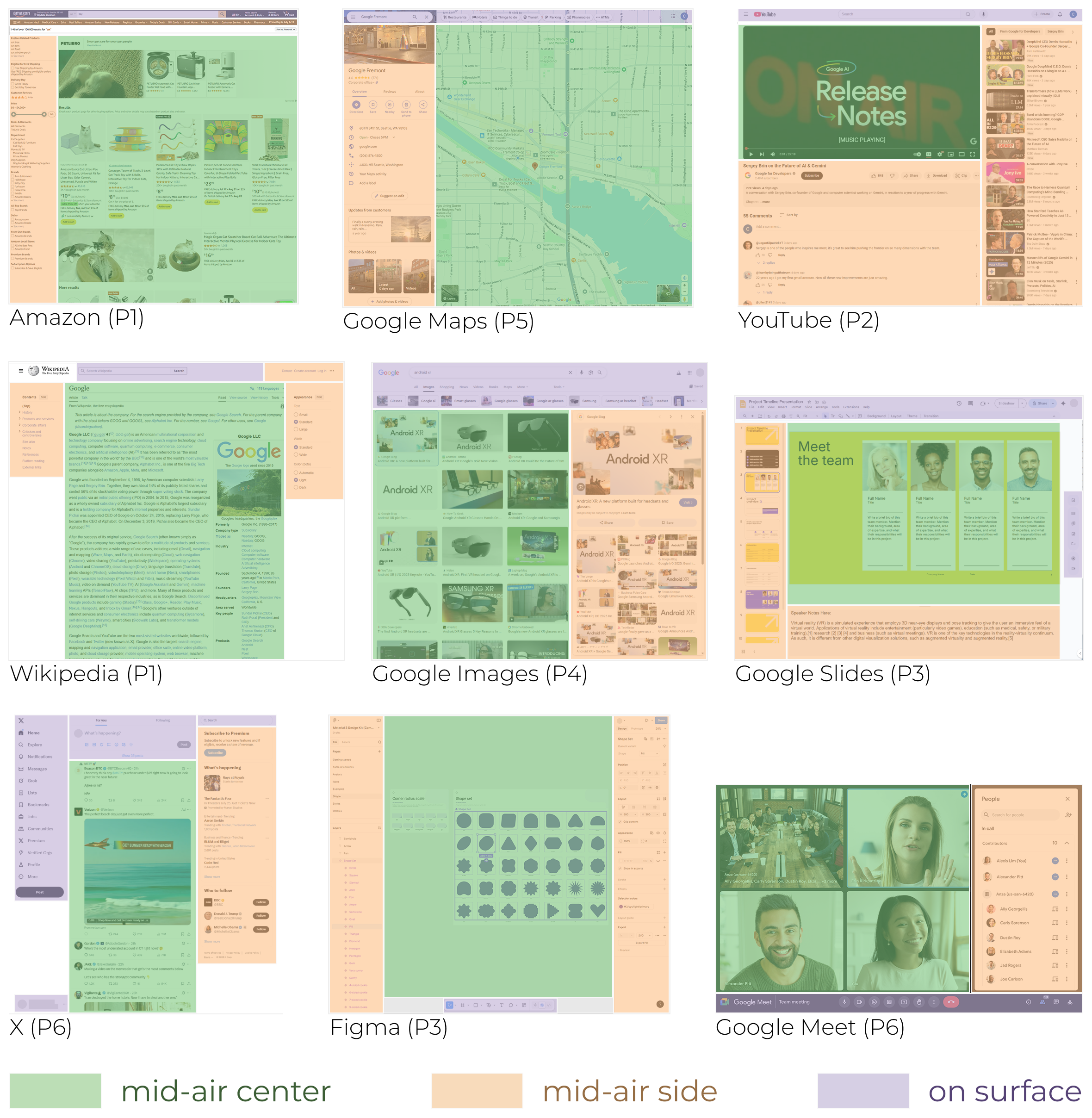}
  \caption{
    Examples of spatial decomposition and placement annotations from our formative study across nine commonly used websites, spanning diverse browsing purposes such as shopping, navigation, media, and content creation.
  }
  \label{fig:formative-study}
\end{figure}

\section{Formative Study and Design Rationales}

To ground our exploration of spatial web browsing, we conducted a formative study examining how XR practitioners decompose and spatially arrange existing webpages. Six experienced XR researchers annotated screenshots from nine commonly used websites, segmenting each page into functional regions and assigning them to spatial locations (e.g., surface-attached, mid-air center, mid-air side) while thinking aloud. Across websites, participants largely preserved familiar webpage structure while reassigning components based on functional role and spatial attention (Figure~\ref{fig:formative-study}), revealing three recurring patterns:

{
\renewcommand{\labelenumi}{R\arabic{enumi}.}

\begin{enumerate}
  \item \textbf{Role-aware decomposition.}  
  Participants treated webpages as collections of functional regions (e.g., primary content, controls, context) rather than indivisible screens, favoring coarse, semantically meaningful chunks.

  \item \textbf{Purposeful spatial roles.}  
  Spatial placement reflected attention and interaction demands: central mid-air space for primary viewing, surfaces for frequent or precise interaction, and peripheral regions for secondary content.

  \item \textbf{Placement--interaction coupling.}  
  Participants expected interaction style to align with spatial placement, associating surfaces with direct touch and mid-air content with indirect input, and often moved content to change how it would be interacted with.
\end{enumerate}
}

\section{Prototype Design: Break-the-Window (BTW)}

We developed \emph{Break-the-Window} (BTW), an exploratory spatial web browsing prototype for Meta Quest~3. BTW decomposes a live webpage into multiple spatial panels that can be placed and interacted with in a mixed-reality workspace, while preserving the original webpage's functionality.

\begin{figure}
    \centering
    \includegraphics[width=\linewidth]{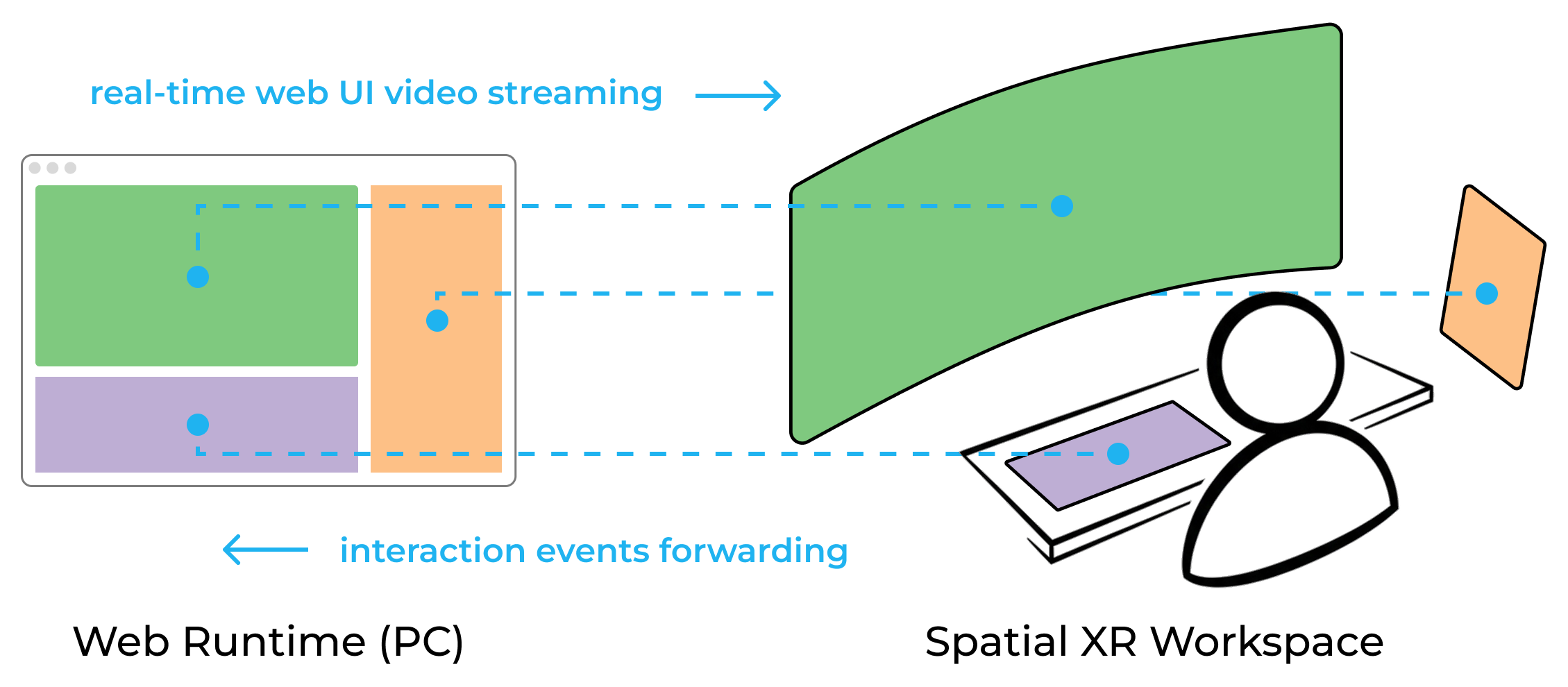}
    \caption{
    Overview of BTW's data flow. A live webpage is rendered in a standard browser and streamed into XR as multiple spatial panels, while user interactions in XR are forwarded back to the original webpage in real time.
    }
    \label{fig:system-overview}
\end{figure}

\subsection{Spatial Web Decomposition and Embodied Interaction}

To support role-aware decomposition without modifying webpage code (R1), BTW adopts a pixel-mirroring approach. An unmodified desktop browser renders a live webpage, whose visual output is streamed into XR. Multiple spatial panels display cropped regions of the same page, and user interactions performed in XR (e.g., clicking, dragging) are relayed back to the original webpage. This allows a single webpage to be decomposed into multiple panels that remain functionally synchronized, enabling BTW to serve as a flexible prototype rather than a fully re-engineered browser.

BTW's frontend supports flexible spatial placement and placement--interaction coupling (R2, R3). Panels can be freely positioned, oriented, scaled, and anchored in mid-air or attached to physical surfaces. Users manipulate panels by grabbing their edges, and panels snap to nearby horizontal surfaces to support desk-based interaction. Interaction is mediated through two complementary techniques: direct touch for nearby panels and ray-based interaction for distant panels. Both techniques support standard web interactions, and transitions between them are handled automatically based on reachability, allowing users to appropriate spatial roles based on task demands rather than fixed layouts.

\begin{figure*}
    \centering

    \begin{subfigure}{0.325\linewidth}
        \centering
        \includegraphics[width=\linewidth]{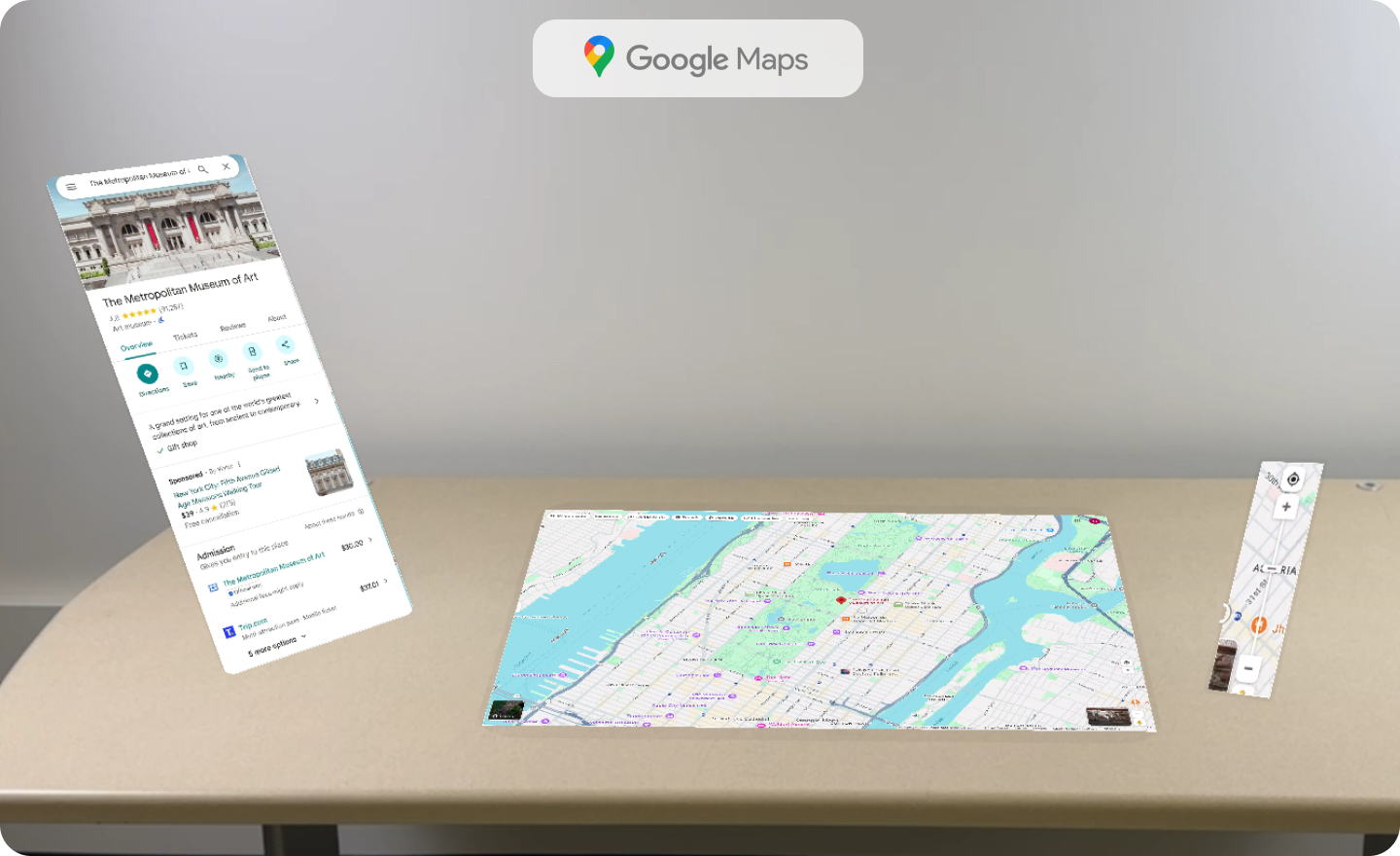}
        \caption{Google Maps}
    \end{subfigure}
    \hfill
    \begin{subfigure}{0.325\linewidth}
        \centering
        \includegraphics[width=\linewidth]{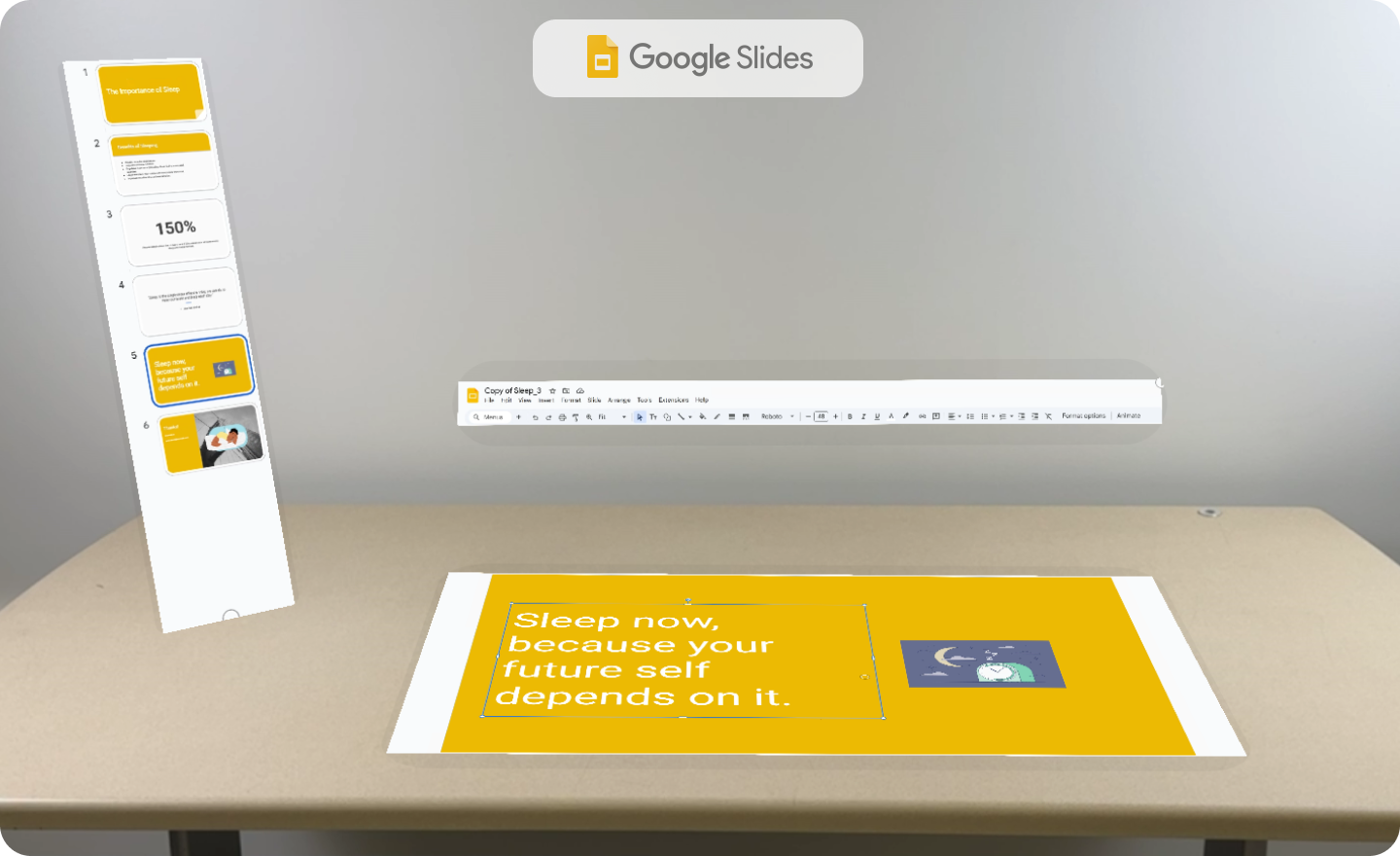}
        \caption{Google Slides}
    \end{subfigure}
    \hfill
    \begin{subfigure}{0.325\linewidth}
        \centering
        \includegraphics[width=\linewidth]{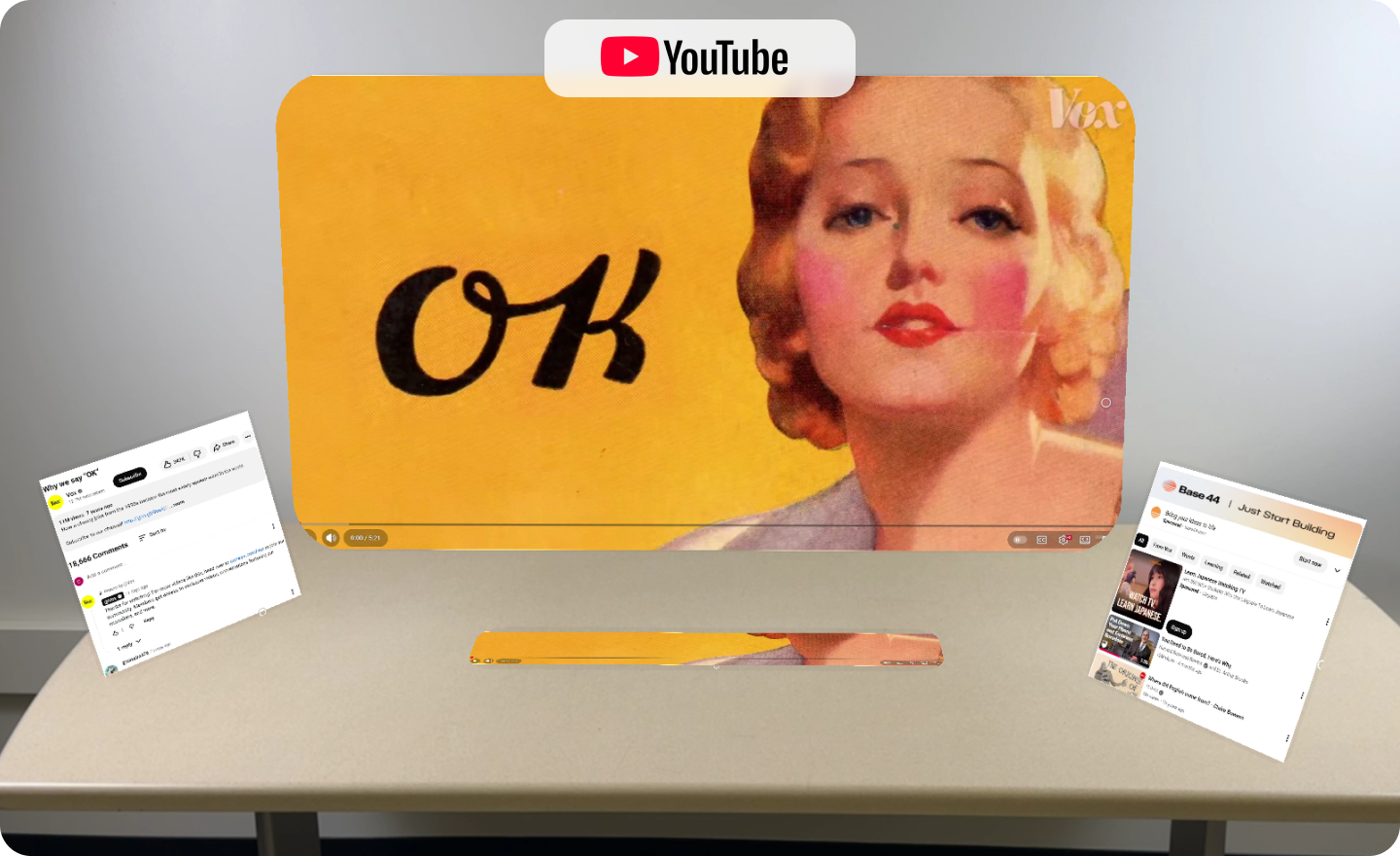}
        \caption{YouTube}
    \end{subfigure}

    \caption{
    Three representative webpages spatialized with \emph{Break-the-Window} (BTW).
    }
    \label{fig:demo-websites}
\end{figure*}

\subsection{Spatial Layouts and Representative Webpages}

BTW provides initial spatial layouts as \emph{design hypotheses} rather than prescriptive configurations. These layouts preserve familiar structural groupings from the original webpage while differentiating spatial roles based on access and interaction needs: interaction-heavy or precision-demanding components are generally placed closer to the user or near the tabletop, while primary or referential content is placed in mid-air. Participants were free to reorganize these layouts during use.

We instantiated BTW with three commonly used webpages --- Google Maps, Google Slides, and YouTube --- to explore how spatial decomposition manifests across different web activities. These examples span continuous versus discrete interaction, viewing versus manipulation, and low versus high precision demands. \textbf{Google Maps} separates the map canvas, information panel, and controls, placing the map near the tabletop to support frequent dragging. \textbf{Google Slides} decomposes the slide canvas, thumbnails, and toolbar, prioritizing precision interaction close to the user. \textbf{YouTube} separates viewing, controls, comments, and recommendations, centering the video player in mid-air while placing interaction-heavy panels closer to the body. Together, these examples illustrate how BTW supports task-sensitive spatial organization without tailoring layouts to a single domain.

\section{User Study}

We conducted an exploratory qualitative study to examine how people interact with and make sense of spatially decomposed web content in XR. The study focused on interaction strategies and spatial organization, rather than performance or efficiency.

\begin{figure*}
    \centering
    \includegraphics[width=\linewidth]{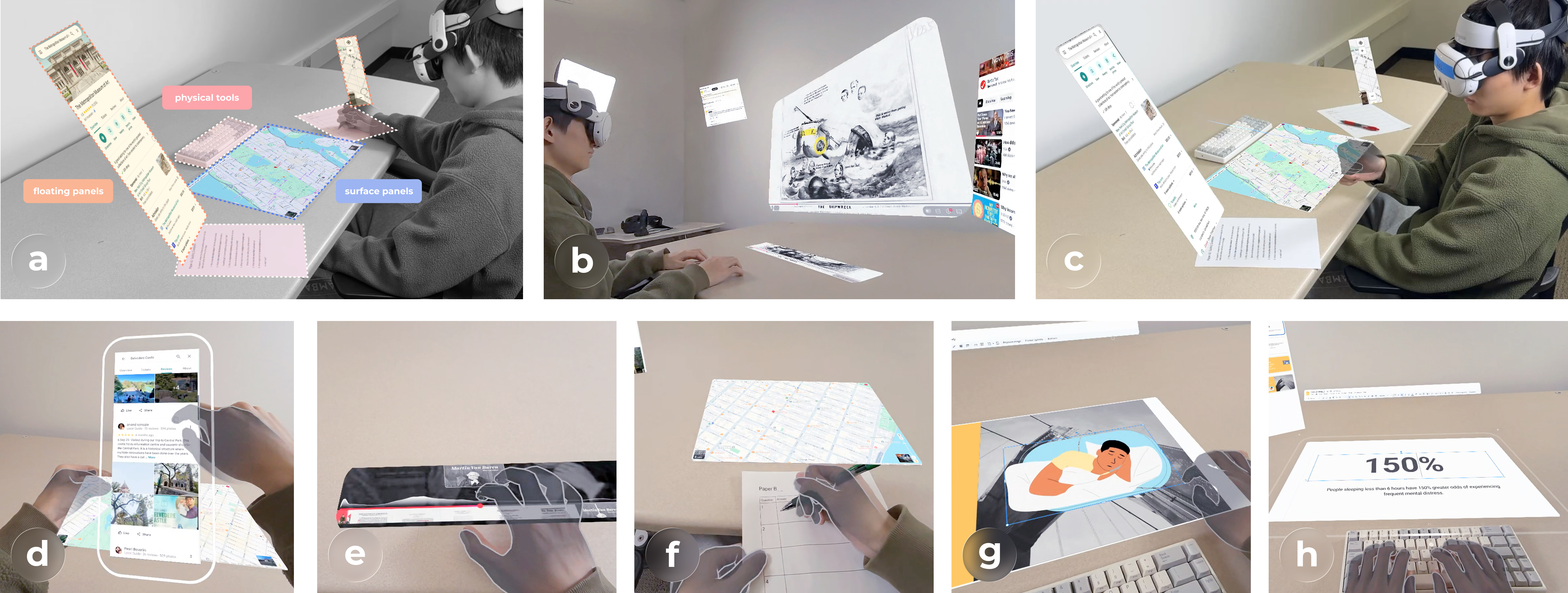}
    \caption{
    A collection of runtime snapshots from the \emph{Break-the-Window} (BTW) system.
    }
    \label{fig:collection}
\end{figure*}

Fifteen participants used \emph{Break-the-Window (BTW)} alongside two experiential reference points---a desktop browser and a single-window XR browser---while completing task-based activities across two of three representative websites (Google Maps, Google Slides, YouTube). The study took place in passthrough VR within a desk workspace, allowing participants to combine spatialized web panels with physical tools such as a keyboard, paper, and pen (Figure~\ref{fig:collection}-A). Following the tasks, participants participated in semi-structured interviews. 
Interview data were analyzed using reflexive thematic analysis to identify recurring patterns in spatial organization, attention, and interaction, which form the basis of the findings reported next. Additional details about the study protocol, tasks, and qualitative analysis process are provided in the supplementary materials.

\section{Key Findings}

Our user study revealed five recurring patterns in how participants organized and interacted with spatially decomposed web content in BTW. Together, these findings show how breaking the browser window transforms browsing into a spatial, embodied workspace---while introducing new interaction frictions.

\noindent\textbf{F1. Attention and interaction become distributed in space.}
Unlike tab switching and scrolling, BTW keeps multiple panels simultaneously visible and spatially separated. Participants used space to maintain peripheral awareness while focusing on a primary view, often describing it as a way to ``block other information'' without closing it (P6). This shift was tightly coupled with embodied navigation, as users reoriented their head, hands, and torso to move between content regions (Fig.~\ref{fig:teaser}-B). At the same time, distributing content increased coordination cost: attending to off-center panels required physical effort, peripheral updates were easier to miss, and repeated turning and reaching accumulated fatigue. Participants adopted distance-dependent input strategies---direct touch for nearby panels and rays for distant or precise actions---though both remained sensitive to small targets and accidental triggers.

\noindent\textbf{F2. Space becomes a semantic substrate.}
Participants treated spatial layout as more than visual arrangement: placement encoded functional role, priority, and relationships (Fig.~\ref{fig:collection}-bc). Near panels were read as ``active'' and interaction-heavy, while farther panels became background. Over time, participants formed location--function bindings (e.g., ``remember the space, not content,'' P9), using spatial memory to retrieve information. However, misplacement proved costly: panels that appeared too near/far, or frequently moved disrupted expectations and led to hesitation.

\noindent\textbf{F3. Physical anchoring both stabilizes and complicates interaction.}
Participants frequently relied on physical metaphors (Fig.~\ref{fig:collection}-de), placing panels on the desk like papers or treating surface panels as ``a virtual iPad'' (P13). Anchoring to real surfaces improved perceived control and precision by providing a stable reference. At the same time, coupling to the physical workspace introduced conflicts: virtual panels occluded paper and tools, writing gestures were misinterpreted as input, and minor surface misalignment undermined trust in touch interaction. (Fig.~\ref{fig:collection}-f)

\noindent\textbf{F4. Immersion-driven engagement does not imply productivity.}
BTW was widely described as novel and engaging, particularly for exploratory and visual scenarios such as browsing and media consumption. However, participants did not consistently feel faster or more efficient. For tool-heavy or precision-oriented work, embodied input slowed throughput: small controls were hard to target, ray interaction demanded careful aiming, and physical effort and attentional demands increased over time (Fig.~\ref{fig:collection}-gh). Participants often contrasted BTW with desktop habits (typing, precise cursor control, rapid switching), describing the desktop as more reliable for routine high-precision tasks.

\noindent\textbf{F5. Spatial web browsing lacks a shared UI grammar.}
Across websites, participants improvised their own layout rules but repeatedly noted the absence of conventions (e.g., ``no rule for where things should go,'' P1). As workspaces grew, this led to clutter, competition with physical space, accidental selections, and fragile mental models. Participants asked for mechanisms that preserve stability (e.g., locking layouts), support legible hierarchy, and provide interaction feedback that aligns with embodied precision.

\section{Discussion: Toward a Spatial UI Grammar for Web Browsing}

Taken together, these findings suggest that once web interfaces are released from a single window, \emph{space becomes part of the interface}. Rather than offering prescriptive guidelines, we outline four structural pressures that point toward emerging design directions for spatial web browsing.

\noindent\textbf{D1. From free placement to interpretable spatial semantics.}
Spatial freedom encourages users to externalize meaning into layout, but also raises the risk of ambiguity and inconsistency as spaces and tasks change. Future spatial web systems may need lightweight structure---e.g., functional grouping, consistent spatial hierarchy, and stable location--function mappings---so that layouts remain legible as they scale.

\noindent\textbf{D2. From static panels to interaction-adaptive representations.}
Web UI elements assume planar viewing and mouse-level precision. In spatial workspaces, distance and posture vary continuously, making fixed sizes and hover-based feedback brittle. Spatial browsers may require distance- and mode-aware representations (e.g., adaptive target sizes/sensitivity, depth-aware feedback, or decoupling precise input from distant viewing).

\noindent\textbf{D3. From spatial expansion to managed spatial capacity.}
Although XR offers more display extent than screens, participants experienced space as limited and costly: larger layouts increased navigation effort and clutter. This motivates mechanisms for managing spatial capacity over time---e.g., layering, collapsing, and transitions between compact and expanded states---while preserving users' mental models.

\noindent\textbf{D4. From interfaces to blended workspaces.}
When panels coexist with desks, paper, and hands, interaction becomes persistent and boundaries blur. Systems should better mediate physical vs.\ digital intent (e.g., through mode switching or intent inference) and leverage environmental constraints (surface structure, body orientation) as stabilizing anchors rather than sources of conflict.

\section{Conclusion}

This poster investigates what changes when webpages are treated not as a single window, but as spatial components embedded in a blended workspace. Through \emph{Break-the-Window (BTW)} and exploratory studies, we saw that spatial decomposition can support distributed attention and meaning-making through layout, and can invite physical metaphors (e.g., desk-based arrangement). At the same time, participants highlighted practical frictions---such as coordination cost across space, precision challenges under embodied input, and conflicts between virtual panels and physical tools---along with a recurring desire for more legible conventions and stability. Rather than proposing a finalized solution, we offer BTW as a concrete research prototype and a starting point for community feedback on how future XR browsers might balance spatial freedom with interpretable structure, interaction adaptivity, and manageable workspace boundaries.

\paragraph{\textbf{Project Availability.}}
The Break-the-Window (BTW) prototype is open-sourced at:
\url{https://github.com/BTW-XR/Break-the-Window}.

\begin{acks}
This material is based upon work supported in part by the National Science Foundation under Grant No. IIS-2441310.
\end{acks}

\bibliographystyle{ACM-Reference-Format}
\bibliography{sample-base,yalong}

@String{Computing = "Computing" }

@inproceedings{hertel2021taxonomy,
  title={A taxonomy of interaction techniques for immersive augmented reality based on an iterative literature review},
  author={Hertel, Julia and Karaosmanoglu, Sukran and Schmidt, Susanne and Br{\"a}ker, Julia and Semmann, Martin and Steinicke, Frank},
  booktitle={2021 IEEE international symposium on mixed and augmented reality (ISMAR)},
  pages={431--440},
  year={2021},
  organization={IEEE}
}

@book{laviola20173d,
  title={3D user interfaces: theory and practice},
  author={LaViola Jr, Joseph J and Kruijff, Ernst and McMahan, Ryan P and Bowman, Doug and Poupyrev, Ivan P},
  year={2017},
  publisher={Addison-Wesley Professional}
}

@article{lu2025ego,
  title={Ego vs. Exo and Active vs. Passive: Investigating the Effects of Viewpoint and Navigation on Spatial Immersion and Understanding in Immersive Storytelling},
  author={Lu, Tao and Zhu, Qian and Ma, Tiffany and Kam-Kwai, Wong and Xie, Anlan and Endert, Alex and Yang, Yalong},
  journal={arXiv preprint arXiv:2502.04542},
  year={2025}
}

@inproceedings{jacob2008reality,
  title={Reality-based interaction: a framework for post-WIMP interfaces},
  author={Jacob, Robert JK and Girouard, Audrey and Hirshfield, Leanne M and Horn, Michael S and Shaer, Orit and Solovey, Erin Treacy and Zigelbaum, Jamie},
  booktitle={Proceedings of the SIGCHI conference on Human factors in computing systems},
  pages={201--210},
  year={2008}
}

@inproceedings{brasier2021ar,
  title={AR-enhanced Widgets for Smartphone-centric Interaction},
  author={Brasier, Eugenie and Pietriga, Emmanuel and Appert, Caroline},
  booktitle={Proceedings of the 23rd International Conference on Mobile Human-Computer Interaction},
  pages={1--12},
  year={2021}
}

@inproceedings{wieland2024push2ar,
  title={Push2ar: enhancing mobile list interactions using augmented reality},
  author={Wieland, Jonathan and Cho, Hyunsung and Hubenschmid, Sebastian and Kiuchi, Akihiro and Reiterer, Harald and Lindlbauer, David},
  booktitle={2024 IEEE International Symposium on Mixed and Augmented Reality (ISMAR)},
  pages={671--680},
  year={2024},
  organization={IEEE}
}

@inproceedings{nebeling2016xdbrowser,
  title={XDBrowser: user-defined cross-device web page designs},
  author={Nebeling, Michael and Dey, Anind K},
  booktitle={Proceedings of the 2016 CHI Conference on Human Factors in Computing Systems},
  pages={5494--5505},
  year={2016}
}

@inproceedings{tan2004wincuts,
  title={WinCuts: manipulating arbitrary window regions for more effective use of screen space},
  author={Tan, Desney S and Meyers, Brian and Czerwinski, Mary},
  booktitle={CHI'04 extended abstracts on Human factors in computing systems},
  pages={1525--1528},
  year={2004}
}

@inproceedings{borowski2025spatialstrates,
  title={Spatialstrates: Cross-reality collaboration through spatial hypermedia},
  author={Borowski, Marcel and Gr{\o}nb{\ae}k, Jens Emil Sloth and Butcher, Peter WS and Ritsos, Panagiotis D and Klokmose, Clemens Nylandsted and Elmqvist, Niklas},
  booktitle={Proceedings of the 38th Annual ACM Symposium on User Interface Software and Technology},
  pages={1--14},
  year={2025}
}

@inproceedings{cheng2023interactionadapt,
  title={Interactionadapt: Interaction-driven workspace adaptation for situated virtual reality environments},
  author={Cheng, Yi Fei and Gebhardt, Christoph and Holz, Christian},
  booktitle={Proceedings of the 36th Annual ACM Symposium on User Interface Software and Technology},
  pages={1--14},
  year={2023}
}

@inproceedings{ball2007move,
  title={Move to improve: promoting physical navigation to increase user performance with large displays},
  author={Ball, Robert and North, Chris and Bowman, Doug A},
  booktitle={Proceedings of the SIGCHI conference on Human factors in computing systems},
  pages={191--200},
  year={2007}
}

@inproceedings{grudin2001partitioning,
  title={Partitioning digital worlds: focal and peripheral awareness in multiple monitor use},
  author={Grudin, Jonathan},
  booktitle={Proceedings of the SIGCHI conference on Human factors in computing systems},
  pages={458--465},
  year={2001}
}

@misc{androidxr2025spatialui,
  author       = {{Android Developers}},
  title        = {Spatial UI},
  howpublished = {\url{https://developer.android.com/design/ui/xr/guides/spatial-ui}},
  year         = {2025},
  note         = {Design \& Plan > UI Design > XR Headsets \& Wired XR Glasses > Guides. Last updated May 16, 2025. Accessed Jan. 2026}
}

@misc{apple2025visionos26,
  author       = {{Apple}},
  title        = {visionOS 26 introduces powerful new spatial experiences for Apple Vision Pro},
  year         = {2025},
  month        = jun,
  day          = {9},
  howpublished = {\url{https://nr.apple.com/DA6k2T2Ax1}},
  note         = {Press release. Accessed Jan. 2026}
}

@inproceedings{luo2026beyond,
  author    = {Luo, Weizhou and Rzayev, Rufat and Russig, Benjamin and 
               Visutarporn, Sivanon and Satkowski, Marc and 
               Gumhold, Stefan and Dachselt, Raimund},
  title     = {Beyond Links: Exploring Visual Representations of Multi-View Relations in Mixed Reality},
  booktitle = {Proceedings of the 2026 CHI Conference on Human Factors in Computing Systems},
  series    = {CHI '26},
  year      = {2026},
  location  = {Barcelona, Spain},
  publisher = {Association for Computing Machinery},
  address   = {New York, NY, USA},
  isbn      = {979-8-4007-2278-3},
  doi       = {10.1145/3772318.3791398},
  url       = {https://doi.org/10.1145/3772318.3791398}
}

@inproceedings{luo2022should,
  title={Where should we put it? layout and placement strategies of documents in augmented reality for collaborative sensemaking},
  author={Luo, Weizhou and Lehmann, Anke and Widengren, Hjalmar and Dachselt, Raimund},
  booktitle={Proceedings of the 2022 CHI Conference on Human Factors in Computing Systems},
  pages={1--16},
  year={2022}
}

@inproceedings{cheng2021semanticadapt,
  title={Semanticadapt: Optimization-based adaptation of mixed reality layouts leveraging virtual-physical semantic connections},
  author={Cheng, Yifei and Yan, Yukang and Yi, Xin and Shi, Yuanchun and Lindlbauer, David},
  booktitle={The 34th Annual ACM Symposium on User Interface Software and Technology},
  pages={282--297},
  year={2021}
}



\end{document}